\documentclass
[superscriptaddress,secnumarabic,amssymb,amsmath,nobibnotes,aps,prd,showkeys,showpacs,nofootinbib,onecolumn]{revtex4}%
\usepackage{graphicx}
\usepackage{subfigure}
\usepackage{epstopdf}
\usepackage{epsf}
\usepackage{bm}
\usepackage{amsmath}
\usepackage{amsfonts}
\usepackage{amssymb}%
\providecommand{\U}[1]{\protect\rule{.1in}{.1in}}

\newcommand{\be}{\begin{equation}}
\newcommand{\ee}{\end{equation}}

\newcommand{\mincir}{\raise
-3.truept\hbox{\rlap{\hbox{$\sim$}}\raise4.truept\hbox{$<$}\ }}
\newcommand{\magcir}{\raise
-3.truept\hbox{\rlap{\hbox{$\sim$}}\raise4.truept\hbox{$>$}\ }}

\usepackage{color}
\begin{document}
\title{Dynamics of Quintessence in Generalized Uncertainty Principle}
\author{Alex Giacomini}
\email{alexgiacomini@uach.cl}
\affiliation{Instituto de Ciencias F\'{\i}sicas y Matem\'{a}ticas, Universidad Austral de
Chile, Valdivia 5090000, Chile}
\author{Genly Leon}
\email{genly.leon@ucn.cl}
\affiliation{Departamento de Matem\'{a}ticas, Universidad Cat\'{o}lica del Norte, Avda.
Angamos 0610, Casilla 1280 Antofagasta, Chile}
\author{Andronikos Paliathanasis}
\email{anpaliat@phys.uoa.gr}
\affiliation{Instituto de Ciencias F\'{\i}sicas y Matem\'{a}ticas, Universidad Austral de
Chile, Valdivia 5090000, Chile}
\affiliation{Institute of Systems Science, Durban University of Technology, Durban 4000,
South Africa}
\author{Supriya Pan}
\email{supriya.maths@presiuniv.ac.in}
\affiliation{Department of Mathematics, Presidency University, 86/1 College Street, Kolkata
700073, India}

\begin{abstract}
We investigate the quintessence scalar field model modified by the Generalized Uncertainty Principle in the background of a spatially flat homogeneous and isotropic universe.  By performing a dynamical system analysis we examine the nature of the critical points and their stability for two potentials, one is the exponential potential and the other is a general potential. In the case of an exponential potential, we find some new critical points for this modified quintessence scenario that describe the de Sitter universes, and these critical points do not appear in the standard quintessence model with an exponential potential. This is one of the main results of this work. Now for the general potential our analysis shows that the physical properties of the critical points remain exactly the same as for the exponential potential which means that within this modified quintessence scenario all kind of potentials have same behaviour. This kind of result is completely new in cosmology because with the change of the potential, differences are usually expected in all respect.  
\end{abstract} 
\keywords{Quintessence; Cosmology; Generalized Uncertainty Principle}
\pacs{98.80.-k, 95.35.+d, 95.36.+x}
\date{\today}
\maketitle

\section{Introduction}

\label{sec1}

The amount of theoretical and observational results strongly suggest that our universe had undergone one accelerated phase of expansion during its very early phase of evolution, known as inflation \cite{guth,linde}, and presently it is again undergoing an accelerating phase of expansion, known as late time cosmic acceleration,  dominated either by some hypothetical dark energy fluid adjusted in the Einstein's General Relativity (GR) \cite{copeland} or by some extra geometrical terms appearing due to  modifications of GR or due to new gravitational theories \cite{Clifton1,Nojiri:2017ncd}. So, one accelerating phase (inflation) occurred in the high energy regime and the second acceleration is occurring in the low energy regime and the difference of the energy scales between the early and late accelerating phases of the universe is significantly high. Therefore, understanding the evolution of the universe in both these regimes has been one of the longstanding issues in modern cosmology. Despite of many investigations performed in these sectors, the actual nature is still unrevealed. The unification of the gravitational theory 
both at classical and quantum levels has therefore remained as an open problem in modern cosmology.

Concerning the early evolution of the universe, considerable attention on the nature of the gravitational theory at the quantum level has been paid by many investigators. The existing literature demands that the theory of quantum gravity \cite{MB01, Niem01, Kowa01, Niem01b, Kempf01, KN01, EGKS01, Amjad1, Amjad2} played a very crucial role to understand the very early evolution of the universe. However, the approach to reveal the quantum nature of the gravity is not unique although. Various approaches were proposed over the years that include  string theory (ST) \cite{Mukhi:2011zz},
doubly special relativity (DSR) \cite{KowalskiGlikman:2004qa,AmelinoCamelia:2010pd,Ghosh:2006bx,Pramanik:2012fj}, black hole physics (BHP) \cite{Bekenstein1, Bekenstein2, Bekenstein3, Bekenstein4} etc. All of them  reported  the existence of a minimum length scale of the order of the Planck length $l_{\rm pl}$ (equivalently, there exists a maximum energy scale in nature). This actually motivated to generalize  the
Heisenberg's Uncertainty Principle to some Generalized Uncertainty principle (GUP) in quantum gravity  \cite{Maggiore}.

The Generalized Uncertainty principle has been found to be very effective to explain several issues related to the dynamics of the universe. For instance, the origin of the magnetic fields with microgauss
strength \cite{SFW86,K94}  which is observed in the intergalactic regimes of the universe acquires some explanation if GUP is taken into account  \cite{Amjad04}. The potential application of GUP can be found in the context of black hole mechanics. Following  the Bekenstein's 
entropy relation \cite{Bekenstein1, Bekenstein2, Bekenstein3, Bekenstein4}
and the Hawking temperature \cite{Hawking74, BD82}, the small black holes in the universe should emit black body radiation. Due to this radiation, these small black holes must loss their mass and consequently becomes hotter.  This emission of radiation  continues until they are completely evaporated. However, the above formulation was based under the assumption of a classical black hole metric together with the consideration that the emitting radiation 
of the black hole was much much smaller (can be ignored) compared to the rest mass energy of the black hole. Now, during this emission process, the black holes naturally become smaller and lighter, and certainly within this length scale,  the assumption of classical metric and the ignorance of the 
radiation seem to be invalid. Thus,  whether the black holes will completely evaporate to either photons, ordinary matter particles or vacuum, or whatever it be (some remnant for instance) is highly questionable due to the lack of a definite quantum gravitational theory (see the discussions \cite{Bunch81, CMP88, York, Parikh99, Susskind95}). Here,  GUP finds an answer towards this discrepancy \cite{Adler}. As argued in Ref. \cite{Adler} a small black 
hole having temperature greater than the ambient temperature will keep  radiating  photons and ordinary particles until it reaches the Planck scale. When the black hole will reach the Planck size, it will stop radiating and its entropy reaches zero while  its effective temperature will reach the maximum limit, and finally, it reduces to an inert 
remnant having no radiating power, but only gravitational interactions may exist. It is not necessary that the created remnant will enjoy the horizon structure as in classical black hole \cite{York}. Such remnants might be consider as the dark matter candidate \cite{Cline}. This is a very surprising result because GUP indicates the origin of one of the heavy fluids of the universe. Naturally, the effects of GUP in various cosmological theories can be studied in order to see whether such effects give rise to some interesting directions related the dynamical history of the universe.  
Since  dark energy is another heavy resource of the universe playing the leading role behind its accelerated expansion, so how GUP modified dark energy models behave, will be an interesting direction of research. As the minimum length effects are quite universal at any stage, so, there is no reason to exclude the GUP modified models, rather, one can generalize such cosmic theories.
Thus, in this work shall focus on the cosmological models that are modified by the GUP. The work has been organized in the following way. 

In Section \ref{sec2} provide with a brief description on the GUP and the corresponding algebra. In Section \ref{sec3} we discuss the quintessence scalar field model modified by the GUP. Then in Section \ref{sec4} we discuss the gravitational field equations of the modified quintessence scalar field in the Friedmann-Lema\^{i}tre-Robertson-Walker universe. 
In Section \ref{sec5} we define the dimensionless dynamical variables and perform the stability analysis. After that in Section \ref{sec6} we introduce the kinematical quantities and compare with the present model. Finally, in Section \ref{sec7} we close the work with a brief summary of all the findings.

\section{Generalized Uncertainty Principle}

\label{sec2}

The GUP is related with the existence of a minimum measurable length, such
that the Heisenberg uncertainty is to be modified as
\begin{equation}
\Delta X_{i}\Delta P_{j}\geqslant\frac{\hbar}{2}[\delta_{ij}(1+\beta
P^{2})+2\beta P_{i}P_{j}]. \label{GUP}%
\end{equation}
where $\beta$ is the deformation parameter defined as $\beta={\beta}_{0}%
\ell_{Pl}^{2}/2\hbar^{2}$, which is also known as quadratic generalized GUP.
From (\ref{GUP}) the generalized deformation of the Heisenberg algebra in a
four-dimensional Minkowski spacetime of signature
\cite{Quesne2006,Vagenas,Kemph1,Kemph2} follows%
\begin{equation}
\lbrack X_{\mu},P_{\nu}]=-i\hbar\lbrack(1-\beta(\eta^{\alpha\gamma}P_{\alpha
}P_{\gamma}))\eta_{\mu\nu}-2\beta P_{\mu}P_{\nu}]. \label{con11} 
\end{equation}
From the algebra (\ref{con11}) it leads to the deformation of the coordinate
representation of the momentum operator \cite{ss1}.

By using the latter commutation relation we can write the coordinate
representation of the operators $X_{\mu},P_{\nu}$. We select $X_{\mu}$
undeformed, that is $X_{\mu}=x_{\mu}$, and the momentum operator to modified
as \cite{Moayedi}
\begin{equation}
P_{\mu}=p_{\mu}(1-\beta(\eta^{\alpha\gamma}p_{\alpha}p_{\gamma}))~,
\label{xp-form}%
\end{equation}
in which~$p^{\mu}$ are defined as usual, that is, $p^{\mu}=i\hbar
\frac{\partial}{\partial x_{\mu}},$ and $[x_{\mu},p_{\nu}]=-i\hbar\eta_{\mu
\nu}$.~

In the context of the GUP, the Klein-Gordon equation which describes a spin-0
particle, described by the wave function $\Psi$, it is now a fourth-order
partial differential equation as given
\begin{equation}
\Delta\Psi+2\beta\hbar\Delta\left(  \Delta\Psi\right)  +\left(  \frac
{mc}{\hbar}\right)  ^{2}\Psi+O\left(  \beta^{2}\right)  =0, \label{kg.01}%
\end{equation}
where $\Delta$ is the Laplace operator defined as $\Delta=\square$ when the
underlying space is $\eta_{\mu\nu}$, or in general $\Delta=\frac{1}{\sqrt{-g}%
}\partial_{\mu}\left(  g^{\mu\nu}\sqrt{-g}\partial_{\mu}\right)  $ for any
Riemannian space of Lorentzian signature with metric $g_{\mu\nu}$. Equation
(\ref{kg.01}) is a singular perturbative equation.

An important observation that helps us to generalize the quintessence scalar
field model in GUP, is that the fourth-order differential equation
(\ref{kg.01}) follows from the variation of the action integral%
\begin{equation}
S_{SF}=\int dx^{4}\sqrt{-g}L\left(  \Psi,\mathcal{D}_{\sigma}\Psi\right),
\label{kg.02}%
\end{equation}
where $L\left(  \Psi,\mathcal{D}_{\sigma}\Psi\right)  $ is the usual
Lagrangian for the Klein-Gordon equation%
\begin{equation}
L\left(  \Psi,\mathcal{D}_{\sigma}\Psi\right)  =\frac{1}{2}g^{\mu\nu
}\mathcal{D}_{\mu}\Psi\mathcal{D}_{\nu}\Psi-\frac{1}{2}\left(  \frac{mc} %
{\hbar}\right)  ^{2}\Psi^{2}, \label{kg.03} %
\end{equation}
where the new operator $\mathcal{D}_{\mu}$ is defined as%
\begin{equation}
\mathcal{D}_{\mu}=\nabla_{\mu}+\beta\hbar^{2}\nabla_{\mu}\left(
\Delta\right)  , \label{kg.04} %
\end{equation}
where $\nabla_{\mu}$ is the covariant derivative for the metric tensor
$g_{\mu\nu}$.

We introduce the new field $\Phi=\Delta\Psi$, and the Lagrange multiplier
$\lambda.$ Variation with respect to the Lagrange multiplier gives,
$\frac{\delta S}{\delta\lambda}=0$ from where it follows $\lambda=-2\beta
\hbar^{2}\Phi$, such that the action integral (\ref{kg.02}) to be written as%
\begin{equation}
S=\int dx^{4}\sqrt{-g}\left(  \frac{1}{2}g^{\mu\nu}\nabla_{\mu}\Psi\nabla
_{\nu}\Psi+2\beta\hbar^{2}g^{\mu\nu}\nabla_{\mu}\Psi\nabla_{\nu}\Phi
+\beta\hbar^{2}\Phi^{2}-\frac{1}{2}V_{0}\Psi^{2}\right)  . \label{pd.66}%
\end{equation}

Therefore, the new Lagrangian is%
\begin{equation}
L\left(  \Psi,\Psi_{;\mu},\Phi,\Phi_{;\mu}\right)  =\sqrt{-g}\left(  \frac
{1}{2}g^{\mu\nu}\Psi_{;\mu}\Psi_{;\nu}+2\beta\hbar^{2}g^{\mu\nu}\Psi_{;\mu
}\Phi_{;\nu}\right)  -\sqrt{-g}\left(  \frac{1}{2}V_{0}\Psi^{2}-\beta\hbar
^{2}\Phi^{2}\right)  \label{pd.67} %
\end{equation}
Hence, the Euler-Lagrange equations 
for Lagrangian (\ref{pd.67})
are%
\begin{equation}
g^{\mu\nu}\Psi_{,\mu\nu}-\Gamma^{\mu}\Psi_{,\mu}-\Phi=0 \label{pd.69}%
\end{equation}%
\begin{equation}
2\beta\hbar^{2}\left(  g^{\mu\nu}\Phi_{,\mu\nu}-\Gamma^{\mu}\Phi_{,\mu
}\right)  +\left(  V_{0}\Psi+\Phi\right)  =0 \label{pd.70}%
\end{equation}
where equation (\ref{pd.69}) is the constraint $\Phi=\Delta_{g}\Psi$, and
system (\ref{pd.69}), (\ref{pd.70}) is equivalent with the fourth-order
partial differential equation (\ref{kg.01}).

\section{Quintessence modified by the GUP}

\label{sec3}

Quintessence is one of the simple dark energy models. Specifically, a canonical
scalar field $\phi\left(  x^{\mu}\right)  $ is responsible for the
acceleration era of the universe. The late time acceleration or the early
acceleration phase, where the scalar field plays the role of the inflaton
\cite{ratra1}.

In the case of a four-dimensional Riemannian manifold described by the metric
tensor $g_{\mu\nu}$, with Ricci scalar $R$, the Action integral which describes
the gravitational theory for the quintessence model is%
\begin{equation}
S_{Q}=\int dx^{4}\sqrt{-g}\left(  -\frac{R}{2}+\frac{1}{2}g^{\mu\nu}%
\nabla_{\mu}\phi\nabla_{\nu}\phi-V(\phi)\right)  . \label{kg.05}%
\end{equation}
However, by using the operator $\mathcal{D}_{\mu}$ defined in (\ref{kg.04}),
we can generalize the latter Action integral as follows%
\begin{equation}
S_{Q}^{GUP}=\int dx^{4}\sqrt{-g}\left(  -\frac{R}{2}+\frac{1}{2}g^{\mu\nu
}\mathcal{D}_{\mu}\phi\mathcal{D}_{\nu}\phi-V(\phi)\right)  , \label{kg.06}%
\end{equation}
which reduces to (\ref{kg.05}) when $\beta=0$.

However, we can always define a new field $\psi=\Delta\phi$, as before and
write the Action Integral (\ref{kg.06}) like that of a second-order theory,
that is,%
\begin{equation}
S_{Q}^{GUP}=\int dx^{4}\sqrt{-g}\left(  -\frac{R}{2}+\frac{1}{2}g^{\mu\nu
}\nabla_{\mu}\phi\nabla_{\nu}\phi-V(\phi)+\beta\hbar^{2}\left(  2g^{\mu\nu
}\nabla_{\mu}\phi\nabla_{\nu}\psi+\psi^{2}\right)  \right)  . \label{kg.09}%
\end{equation}

Variation with respect to the metric tensor of (\ref{kg.09}) leads to the
gravitational field equations, which are,
\begin{equation}
G_{\mu\nu}=T_{\mu\nu}, \label{kg.10}%
\end{equation}
where $G_{\mu\nu}$ is the Einstein tensor and~$T_{\mu\nu}$ is the energy
momentum tensor for the two fields defined as
\begin{align}
T_{\mu\nu}  &  =\left(  \frac{1}{2}\nabla_{\mu}\phi\nabla_{\nu}\phi
+2\beta\hbar^{2}\nabla_{\mu}\phi\nabla_{\nu}\psi\right)  +\nonumber\\
&  -g_{\mu\nu}\left(  \frac{1}{2}g^{\alpha\beta}\nabla_{\alpha}\phi
\nabla_{\beta}\phi+2\beta\hbar^{2}g^{\alpha\beta}\nabla_{\alpha}\phi
\nabla_{\beta}\psi-V(\phi)+\beta\hbar^{2}\psi^{2}\right)  , \label{kg.11}%
\end{align}

Furthermore, variation of (\ref{kg.09}) with respect to the scalar fields $\phi,~\psi$ leads to
the equation of motions for the two fields%
\begin{equation}
\Delta\phi-\psi=0, \label{kg.12}%
\end{equation}%
\begin{equation}
\beta\hbar^{2}\Delta\psi+\frac{1}{2}\left(  \psi+V_{,\phi}\right)  =0.
\label{kg.13}%
\end{equation}

\section{FLRW background space}

\label{sec4}

In the large scales our universe is almost homogeneous and isotropic and
such description of the universe is well described by the Friedmann-Lema\^{i}tre-Robertson-Walker (FLRW) line element with zero spatial curvature given by,
\begin{equation}
ds^{2}=dt^{2}-a^{2}\left(  t\right)  \left(  dx^{2}+dy^{2}+dz^{2}\right)  ,
\label{frw.01}%
\end{equation}
where $a\left(  t\right)  $ is the expansion scale factor of the universe, and $H=\dot{a}/a$ is
the Hubble function.

For the line element (\ref{frw.01}) the gravitational field equations
(\ref{kg.10}) are%
\begin{equation}
3H^{2}=\left(  \frac{1}{2}\dot{\phi}^{2}+V\left(  \phi\right)  \right)
+\beta\hbar^{2}\left(  2\dot{\phi}\dot{\psi}-\psi^{2}\right)  +\rho_{m}
\label{frw.02}%
\end{equation}%
\begin{equation}
2\dot{H}+3H^{2}=-\left(  \frac{1}{2}\dot{\phi}^{2}-V\left(  \phi\right)
+\beta\hbar^{2}\left(  2\dot{\phi}\dot{\psi}+\psi^{2}\right)  \right)  ,
\label{frw.03}%
\end{equation}
where $\rho_{m}$ is the energy density for the dust fluid source, i.e.
$T_{\mu\nu}^{(m)}=\rho_{m}u_{\mu}u_{\nu}$, in which $u^{\mu}=\delta_{t}^{\mu}$.

For the fields $\phi,~\psi$ the equation of motions are%
\begin{align}
\ddot{\phi}+\frac{3}{a}\dot{a}\dot{\phi}-\psi &  =0,\label{frw.04}\\
\varepsilon\left(  \ddot{\psi}+\frac{3}{a}\dot{a}\dot{\psi}\right)  +\frac
{1}{2}\left(  \psi+V_{,\phi}\right)   &  =0, \label{frw.05}%
\end{align}
while for the dust fluid source, the Bianchi identity $T_{~\ ~~~~;\nu
}^{(m)~\mu\nu}=0$ gives $\dot{\rho}_{m}+3H\rho_{m}=0,$ that is $\rho_{m}%
=\rho_{m0}a^{-3}$, where $\rho_{m0}$ is the present value of $\rho_m$. 

An equivalent way to write the field equations is to define the new variables%
\begin{align}
&  3H^{2}=\rho_{GUP}+\rho_{m},\\
&  2\dot{H}+3H^{2}=-p_{GUP},
\end{align}
where
\[
\rho_{GUP}=\rho_{\phi}+\rho_{\psi},~p_{GUP}=p_{\phi}+p_{\psi}.
\]

Parameters $\rho_{\phi}$, $~p_{\phi}$ are the energy density and pressure for
the quintessence
\begin{equation}
\rho_{\phi}=\frac{1}{2}\dot{\phi}^{2}+V\left(  \phi\right)  ~,~p_{\phi}%
=\frac{1}{2}\dot{\phi}^{2}-V\left(  \phi\right)  ,
\end{equation}
and $\rho_{\psi}$, $p_{\psi}$ are the energy and pressure for the second
interacting field $\psi$, that is,
\begin{equation}
\rho_{\psi}=\beta\hbar^{2}\left(  2\dot{\phi}\dot{\psi}-\psi^{2}\right)
~,~p_{\psi}=\beta\hbar^{2}\left(  2\dot{\phi}\dot{\psi}+\psi^{2}\right)  .
\end{equation}
At this point we remark that the second field $\psi$ is not a physical field,
but it is introduced by the quantum corrections of GUP and it describes the
new geometrodynamic degrees of freedom of GUP.

As far as concerns the parameter for the equation of state of the scalar
field, that is calculated
\begin{equation}
w_{GUP}=\frac{p_{GUP}}{\rho_{GUP}}=\frac{\left(  \frac{1}{2}\dot{\phi}%
^{2}-V\left(  \phi\right)  \right)  +\beta\hbar^{2}\left(  2\dot{\phi}%
\dot{\psi}+\psi^{2}\right)  }{\left(  \frac{1}{2}\dot{\phi}^{2}+V\left(
\phi\right)  \right)  +\beta\hbar^{2}\left(  2\dot{\phi}\dot{\psi}-\psi
^{2}\right)  }. \label{ac.20}%
\end{equation}
where expanding around~$\beta\hbar^{2}\rightarrow0$, it follows
\begin{equation}
w_{GUP}\simeq w_{\phi}+\frac{1}{\rho_{\phi}}\left(  p_{\psi}-w_{\phi}%
\rho_{\psi}\right)  +O\left(  \left(  \beta\hbar^{2}\right)  ^{2}\right)
\label{ac.20c}%
\end{equation}
while when the scalar field potential $V\left(  \phi\right)  $, dominates, that
is $\dot{\phi}^{2}\ll V$ it follows%
\begin{equation}
w_{GUP}\simeq-1+\frac{4\beta\hbar^{2}}{V}\dot{\phi}\dot{\psi}%
\end{equation}
which can cross the phantom divide line when $~\dot{\phi}\dot{\psi}\leq0$. An
important observation here is that the field equations form a singular
pertubation system, also known as slow-fast dynamical system. That means that
it is possible$~\left\vert \dot{\psi}\right\vert $ to be large enough such
that the term $\left\vert \beta\hbar^{2}\dot{\psi}\right\vert \ $\ will not be
negligible. That is a main characteristic of the slow-fast dynamical systems.

In the following sections we study the global evolution of the dynamics, in
order to understand the effects of GUP in the quintessence field. This work
extends and generalize the previous work on the specific theory \cite{Paliathanasis:2015cza}.

\section{Dynamical systems formulation}

\label{sec5}

We continue our analysis by define the new dimensionless variables%
\begin{equation}
x_{1}=\frac{\dot{\phi}}{\sqrt{6}H}~,~y_{1}=\sqrt{\frac{V}{3H^{2}}}%
~,~x_{2}=\beta\hbar^{2}\frac{2\sqrt{2}\dot{\psi}}{\sqrt{3}H}~,~y_{2}%
=\frac{\beta\hbar^{2}\psi^{2}}{3H^{2}},~\Omega_{m}=\frac{\rho_{m}}{3H^{2}%
},\label{gp.01}%
\end{equation}
such that the gravitational field equations to be written as%
\begin{equation}
\frac{dx_{1}}{d\tau}=\frac{1}{4}\left(  6x_{1}\left(  x_{1}\left(  x_{1}%
+x_{2}\right)  -y_{1}^{2}-1\right)  +y_{2}\left(  6x_{1}-\sqrt{6}\mu\right)
\right)  ,\label{gp.02}%
\end{equation}%
\begin{equation}
\frac{dy_{1}}{d\tau}=\frac{1}{2}y_{1}\left(  3\left(  1-y_{1}^{2}%
+y_{2}\right)  +x_{1}\left(  3\left(  x_{1}+x_{2}\right)  -\sqrt{6}%
\lambda\right)  \right)  ,\label{gp.03}%
\end{equation}%
\begin{equation}
\frac{dx_{2}}{d\tau}=\frac{1}{2}\left(  3x_{1}x_{2}^{2}-3x_{2}\left(
1-x_{1}^{2}+y_{1}^{2}-y_{2}\right)  +\sqrt{6}\left(  2\lambda y_{1}^{2}+\mu
y_{2}\right)  \right)  \label{gp.04}%
\end{equation}%
\begin{equation}
\frac{dy_{2}}{d\tau}=\frac{1}{4}y_{2}\left(  12x_{1}\left(  x_{1}%
+x_{2}\right)  +12\left(  1-y_{1}^{2}-y_{2}\right)  -\sqrt{6}x_{2}\mu\right)
\label{gp.05}%
\end{equation}%
\begin{equation}
\frac{d\mu}{d\tau}=\frac{1}{4}\sqrt{\frac{3}{2}}x_{2}\mu^{2}~,~\frac{d\lambda
}{d\tau}=\sqrt{6}x_{1}\lambda^{2}\left(  \Gamma\left(  \lambda\right)
-1\right)  \label{gp.06}%
\end{equation}
where the new independent variable $\tau=\ln a$, and variables $\lambda,\mu$
are defined as%
\begin{equation}
\lambda=-\frac{V_{,\phi}}{V}~,~\beta\hbar^{2}\mu=-\frac{2}{\psi}\label{gp.07}%
\end{equation}
while $\Gamma\left(  \lambda\right)  =\frac{V_{,\phi\phi}V}{V_{,\phi}^{2}}$.
Moreover, the constraint equation is written as%
\begin{equation}
\Omega_{m}\left(  \mathbf{x,y}\right)  =1-x_{1}\left(  x_{1}+x_{2}\right)
-y_{1}^{2}+y_{2},\label{gp.08}%
\end{equation}
from where it follows that the paramters are constraint as $\,0\leq\Omega
_{m}\leq1.$

The parameter for the equation of state for the effective fluid is derived to
be%
\begin{equation}
w_{tot}\left(  \mathbf{x,y}\right)  =x_{1}\left(  x_{1}+x_{2}\right)
-y_{1}^{2}+y_{2}, \label{gp.09}%
\end{equation}
while we define the physical variables $\Omega_{\phi},~\Omega_{\psi}$ such as,
$\Omega_{\phi}\left(  \mathbf{x,y}\right)  =x_{1}^{2}+y_{1}^{2}~,~\Omega
_{\psi}\left(  \mathbf{x,y}\right)  =x_{1}x_{2}-y_{2},~$where the constraint
equation is written as $\Omega_{m}\left(  \mathbf{x,y}\right)  =1-\Omega
_{\phi}\left(  \mathbf{x,y}\right)  -\Omega_{\psi}\left(  \mathbf{x,y}\right)
$.

Finally, the parameter for the equation of state for the scalar field is
expressed in terms of the new variables as
\begin{equation}
w_{\phi}\left(  \mathbf{x,y}\right)  =\frac{\left(  x_{1}^{2}-y_{1}%
^{2}\right)  +x_{1}x_{2}+y_{2}}{\left(  x_{1}^{2}+y_{1}^{2}\right)
+x_{1}x_{2}-y_{2}}. \label{gp.10}%
\end{equation}

\subsection{Equilibrium points for exponential potential}

Consider now that the scalar field potential is exponential, that is $V\left(
\phi\right)  =V_{0}e^{-\lambda\phi}$, such that $\Gamma\left(  \lambda\right)
=1$ and $\frac{d\lambda}{d\tau}=0$. For that potential the dynamical system
(\ref{gp.02})-(\ref{gp.06}) is reduced by one-dimension. Therefore, the
equilibrium points of the dynamical system are the points with coordinates
$P=\left(  x_{1}\left(  P\right), x_{2}\left(  P\right), y_{1}\left(P\right), y_{2}\left(  P\right), \mu\left(  P\right)  \right) $, where the
right hand side of (\ref{gp.02})-(\ref{gp.06}) are zero.

The equilibrium points are calculated to be%
\begin{equation}
P_{0}=\left(  0,0,0,0,0,\mu\right)  ~,~P_{1}^{\pm}=\left(  \pm1,0,0,0,\mu
\right)  ~,
\end{equation}%
\begin{equation}
~P_{2}=\left(  x_{1},\frac{1}{x_{1}}-x_{1},0,0,0\right)  ~,~P_{3}=\left(
0,\sqrt{\frac{2}{3}}\lambda y_{1}^{2},y_{1},y_{1}^{2}-1,0\right)  .
\end{equation}
In the following lines we discuss the physical properties of the points as
also their stability.

Point $P_{0}\,\ $describes a universe dominated by the dust fluid source,
where $\Omega_{m}\left(  P_{0}\right)  =1$, while the parameter for the
equation of state $w_{tot}$ is derived to be $w_{tot}\left(  P_{0}\right)
=0$. The eigenvalues of the linearized system are calculated to be
$e_{1}\left(  P_{0}\right)  =3~,~e_{2}\left(  P_{0}\right)  =-\frac{3}%
{2}~,~e_{3}\left(  P_{0}\right)  =-\frac{3}{2},~e_{4}\left(  P_{0}\right)
=\frac{3}{2}$ and $e_{5}\left(  P_{0}\right)  =0$, from where we infer that
the point is a saddle point and the exact solution at the point is always unstable.

Poins $P_{1}^{\pm}$ describe universes dominated by the kinetic term of the
scalar field, i.e. $\Omega_{m}\left(  P_{1}^{\pm}\right)  =0$, and
$w_{tot}\left(  P_{1}^{\pm}\right)  =1.$ The eigenvalues are calculated to be
$e_{1}\left(  P_{1}^{\pm}\right)  =6~,~e_{2}\left(  P_{1}^{\pm}\right)
=3~,~e_{3}\left(  P_{1}^{\pm}\right)  =\frac{1}{2}\left(  6\mp\sqrt{6}%
\lambda\right)  ,~e_{4}\left(  P_{1}^{\pm}\right)  =0$ and $e_{5}\left(
P_{1}^{\pm}\right)  =0$, from where we infer that the exact solutions at the
equilibrium points are always unstable. Point $P_{1}^{+}$ is a saddle point
when $\lambda>\sqrt{6}$ while point $P_{1}^{-}$ is a saddle point when
$\lambda<-\sqrt{6}$.

Point $P_{2}$ is actually family of points that lie on a two-dimensional surface.
The physical properties of this points are similar with that of points
$P_{1}^{\pm}$. The eigenvalues of the linearized system are derived
$e_{1}\left(  P_{2}\right)  =6~,~e_{2}\left(  P_{2}\right)  =3~,~e_{3}\left(
P_{2}\right)  =\frac{1}{2}\left(  6-\sqrt{6}x_{1}\lambda\right)
,~e_{4}\left(  P_{2}\right)  =0$ and $e_{5}\left(  P_{2}\right)  =0$, from
where we infer that the exact solutions at the family of points are always unstable.

The family of points on the surface with coordinates $P_{3}$ describe de
Sitter universes where $\Omega_{m}\left(  P_{3}\right)  =0$ and $w_{tot}%
\left(  P_{3}\right)  =-1$. The eigevalues of the linearized system are
determined to be $e_{1}\left(  P_{3}\right)  =-3~,~e_{2}\left(  P_{3}\right)
=-3~,~e_{3}\left(  P_{3}\right)  =-3,~e_{4}\left(  P_{3}\right)  =0$ and
$e_{5}\left(  P_{2}\right)  =0$, hence the Center Manifold Theorem (CMT) should be applied to find the
manifold where the de Sitter universes are attractors.

We observe that no scaling solutions or a tracking solutions exist in this
specific model like in the quintessence theory. However, the critical points
which describe the de Sitter solution do not exist in the case of quintessence
for the exponential potential; these are the new equilibrium points provided by
the new terms given by GUP.

\begin{figure}[ptb]
\centering\includegraphics[width=0.8\textwidth]{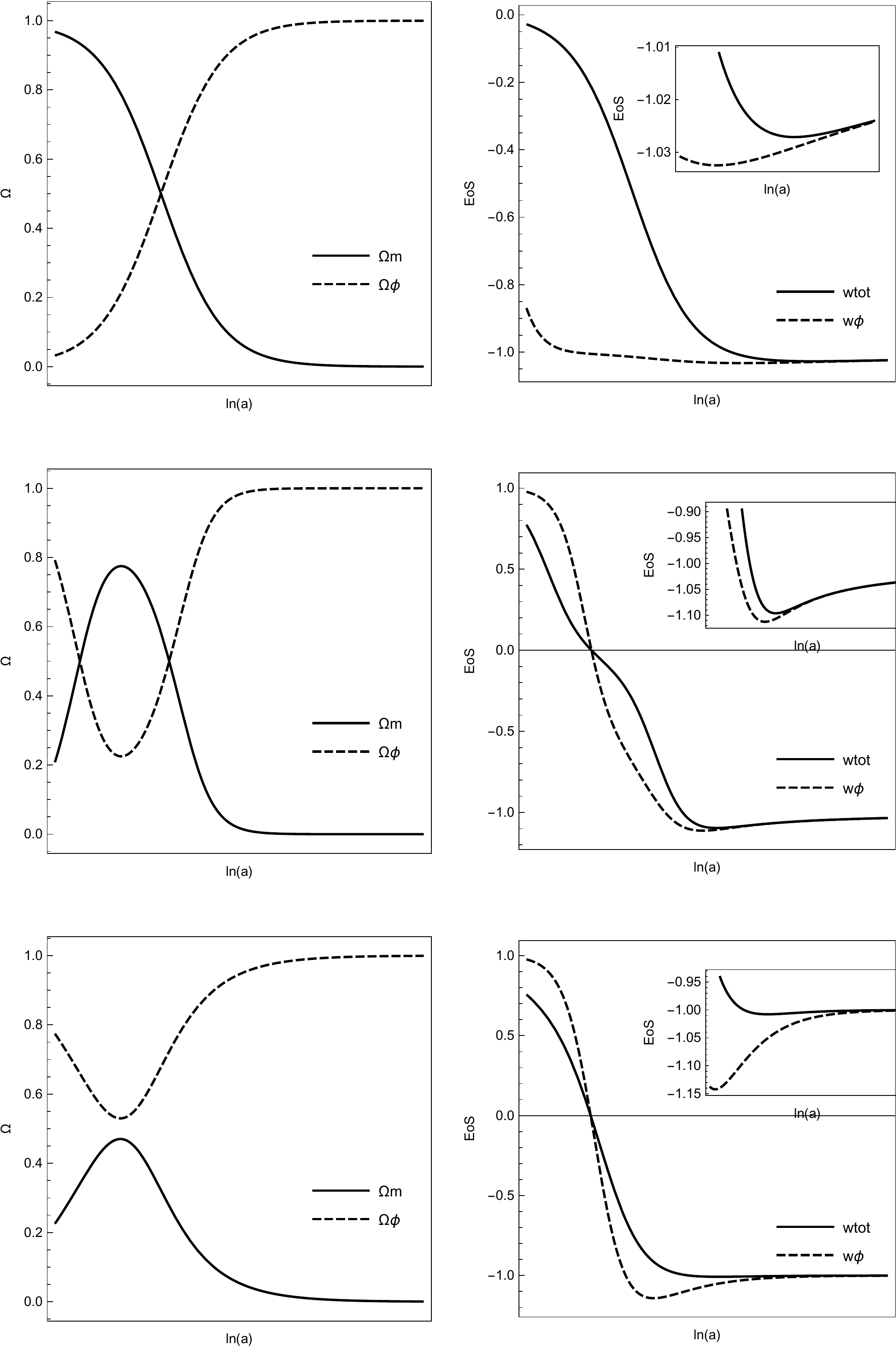} \caption{Qualitative
evolution of the physical variables $\Omega_{m},~\Omega_{\phi}$ (Left
figs.)\ and $w_{tot},~w_{\phi}$ (Right figs) for three sets of initial
conditions. The plots of the first row is for initial conditions near to the
point $P_{0}$, the second row is for initial condition near to the point
$P_{1}^{+}$, while the plots of the third row are for initial conditions near
to the point $P_{1}^{-}$. For the energy density we observe that at late times
the scalar field dominates $\Omega_{\phi}\rightarrow1$ and $\Omega
_{m}\rightarrow0$, while the parameter for the equation of state $w_{tot}$
have the limits $w_{tot}\rightarrow-1$. Thus, we observe that $w_{\phi}$ and
$w_{tot}$ can cross the phantom divide line and take values smaller than minus
one. }%
\label{fig0}%
\end{figure}

We solve numerically the field equations (\ref{gp.02})-(\ref{gp.06}) for three
different sets of initial conditions, and we present the evolution of the
physical parameters $\Omega_{m},~\Omega_{\phi}=1-\Omega_{m}$ and
$w_{tot},~w_{\phi}$ if Fig. \ref{fig0}. For the three different sets of the
initial conditons the final attractor is the de Sitter universe, while for the
parameters of the equation of state we observe that they can cross the phantom
divide line. In order to understand that behaviour we present the phase space
diagram for the dynamical system in the plane $\left\{  x_{1}-y_{1}\right\}  $
from where it is clear that $P_{3}$ can be a local attractor. 

To analyze $P_{3}$ in the following we proceed with
the application of the CMT. 

\begin{figure}[ptb]
\centering\includegraphics[width=0.5\textwidth]{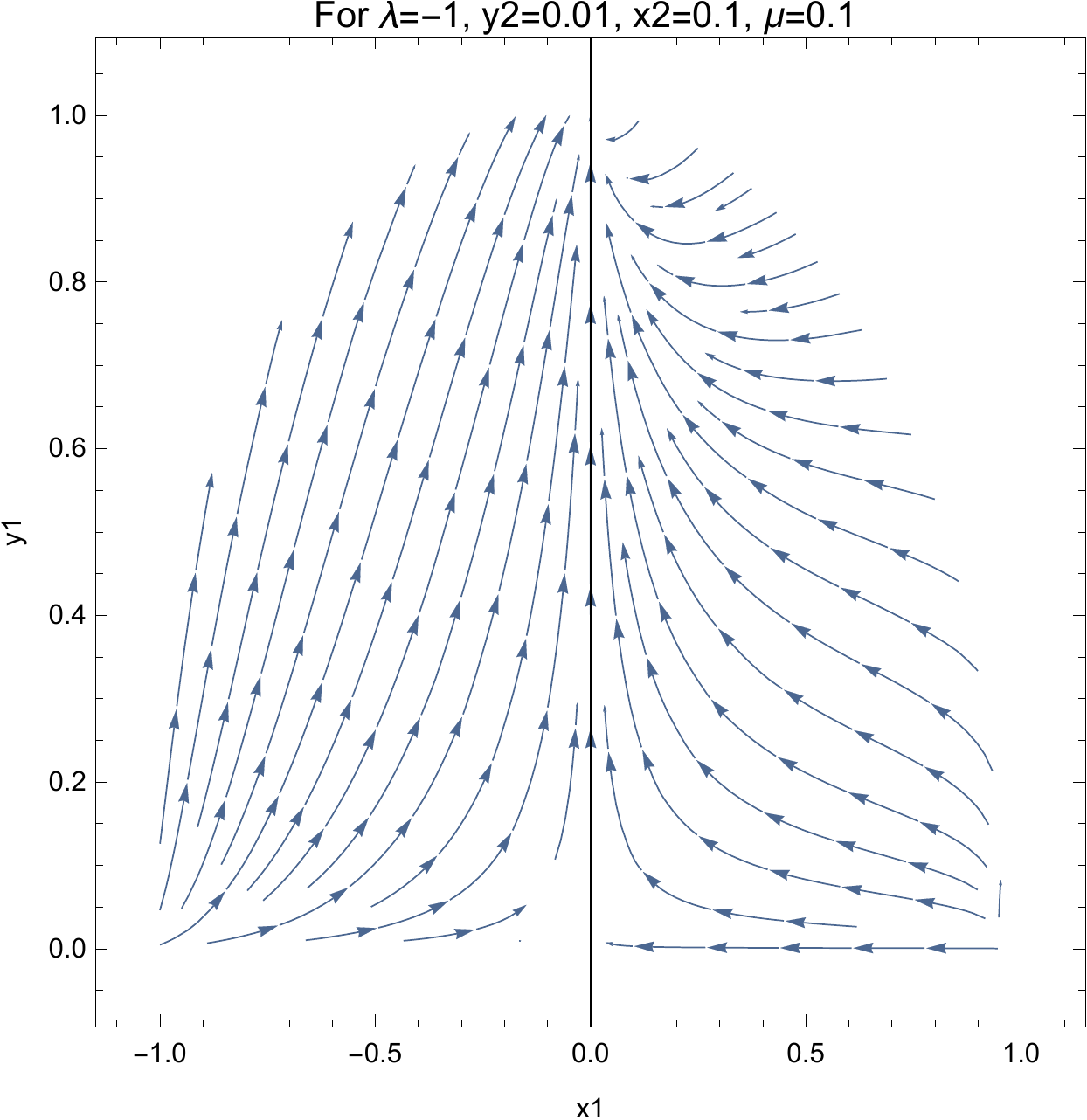} \caption{Phase space
diagram in the plane $\left\{  x_{1}-y_{1}\right\}  $ where for the rest of
the parameters we selected $\lambda=-1$, $y_{2}=0.01$, $x_{2}=0.01$ and
$\mu=0.1$.~From the diagram we observe the attractor $P_{3}.$}%
\label{fig1}%
\end{figure}

\subsubsection{Center manifold theorem  for line of  points $P_{3}$}

Assuming $y_{1c}\notin\left\{  0, \frac{\sqrt{2}}{2}, 1\right\}, \lambda \neq 0$, and introducing the new variables 
\begin{align}
  u_1 &= \frac{1}{3} {y_{1c}} \left(\left({y_{1c}}^2-1\right) \left({y_{1c}} \left(\sqrt{6} \lambda  {x_1}+\lambda  \mu  \left({y_{1c}}^2-1\right)+3\right)-6 {y_1}\right)+3 {y_{1c}}
   {y_2}\right),\\
  u_2 &= -\frac{1}{2} \lambda  \mu  {y_{1c}}^2 \left(2 {y_{1c}}^4-3 {y_{1c}}^2+1\right),\\
   &v_1= \frac{1}{18} \Big\{6 \lambda  {y_{1c}}^2 \left(-2 \lambda  {x_1}
   \left({y_{1c}}^2-1\right)-\sqrt{6} \left({y_{1c}}^2+{y_2}\right)\right)+18 {x_2}+12 \sqrt{6} \lambda  {y_1} \left({y_{1c}}^2-1\right) {y_{1c}} \nonumber \\
   & -\sqrt{6} \mu  \left({y_{1c}}^2-1\right) \left(\lambda ^2
   {y_{1c}}^2 \left(3 {y_{1c}}^2-2\right)+3\right)\Big\},\\
  v_2 &= \frac{1}{6} \lambda  {y_{1c}}^2 \left(-\lambda  \left(6 {x_1} \left({y_{1c}}^2-2\right)+\sqrt{6} \mu  \left({y_{1c}}^2-1\right)^2\right)-3
   \sqrt{6} \left(-2 {y_1} {y_{1c}}+{y_{1c}}^2+{y_2}+1\right)\right),\\
   v_3 & = -\frac{1}{3} \left({y_{1c}}^2-1\right) \left({y_{1c}}^2 \left(\sqrt{6} \lambda  {x_1}+\lambda  \mu 
   \left({y_{1c}}^2-1\right)+3\right)-6 {y_1} {y_{1c}}+3 ({y_2}+1)\right),
\end{align}
a particular  point $P_{3c}=\left(
0,\sqrt{\frac{2}{3}}\lambda y_{1c}^{2},y_{1c},y_{1c}^{2}-1,0\right)$ on line of points $P_3$ is translated to the origin and the linearization of the system \eqref{gp.01}, \eqref{gp.02}, \eqref{gp.03}, \eqref{gp.04}, \eqref{gp.05} is transformed to its real Jordan canonical form:
\begin{equation}
 \left(
\begin{array}{c}
 v_1'\\
 v_2'\\
v_3'\\
u_1'\\
u_2'\\
\end{array}
\right)   =  \left(
\begin{array}{ccccc}
 -3 & 0 & 0 & 0 & 0 \\
 0 & -3 & 1 & 0 & 0 \\
 0 & 0 & -3 & 0 & 0 \\
 0 & 0 & 0 & 0 & 1 \\
 0 & 0 & 0 & 0 & 0 \\
\end{array}
\right)  \left(
\begin{array}{c}
 v_1\\
 v_2\\
v_3\\
u_1\\
u_2\\
\end{array}
\right).
\end{equation}

Assuming $y_{1c}\notin\left\{ 0, \frac{1}{\sqrt{2}}, 1\right\}, \lambda \neq 0$, and applying the CMT we obtain that the center manifold is given locally (up to second order) by the graph
\begin{align}
   \Bigg\{(u_1,u_2, v_1, v_2, v_3)\in\mathbb{R}^5: \quad   & v_1 = a_1 u_1^2 + a_2 u_1 u_2 + a_3 u_2^2 + \mathcal{O}(3), \nonumber \\ 
    & v_2 = b_1 u_1^2 + b_2 u_1 u_2 + b_3 u_2^2 + \mathcal{O}(3), \nonumber \\  
    & v_3 = c_1 u_1^2 + c_2 u_1 u_2 + c_3 u_2^2 + \mathcal{O}(3)\Bigg\},
\end{align}
where
\begin{align}
& a_1= -\frac{\lambda  \left({y_{1c}}^2-1\right)}{2 \sqrt{6} {y_{1c}}^2}, a_2= \frac{\lambda ^2 \left(-11 {y_{1c}}^4+10 {y_{1c}}^2-1\right)-6}{3 \sqrt{6} \lambda  {y_{1c}}^2 \left(2 {y_{1c}}^4-3
   {y_{1c}}^2+1\right)}, a_3=\frac{6 \left(4 {y_{1c}}^2-3\right)+\lambda ^2 \left(40 {y_{1c}}^6-57 {y_{1c}}^4+20 {y_{1c}}^2-1\right)}{18 \sqrt{6} \lambda  \left({y_{1c}}^2-1\right) \left({y_{1c}}-2
   {y_{1c}}^3\right)^2},\\
   & b_1=-\frac{1}{4} \sqrt{\frac{3}{2}} \lambda , b_2=\frac{5 \lambda }{\sqrt{6} \left(2-4 {y_{1c}}^2\right)}, b_3= \frac{\lambda ^2 {y_{1c}}^2 \left(12 {y_{1c}}^2-5\right)+6}{12 \sqrt{6} \lambda  \left({y_{1c}}-2
   {y_{1c}}^3\right)^2},\\
   & c_1= \frac{1}{4} \left(\frac{1}{{y_{1c}}^2}-1\right), c_2=\frac{1}{2-4 {y_{1c}}^2}-\frac{1}{6 {y_{1c}}^2}, c_3= \frac{{y_{1c}}^2 \left(\lambda ^2 \left(12 {y_{1c}}^4-9 {y_{1c}}^2+1\right)+6\right)-6}{36
   {y_{1c}}^4 \left(\lambda -2 \lambda  {y_{1c}}^2\right)^2}.
\end{align}
Hence, the dynamics on the center manifold is given locally (up to second order) by 
\begin{align}
   &u_1'=\frac{{u_1} {u_2} \left(2 {y_{1c}}^6+7 {y_{1c}}^4-8 {y_{1c}}^2+1\right)}{4 {y_{1c}}^6-6 {y_{1c}}^4+2 {y_{1c}}^2}+\frac{{u_2}^2 \left(\lambda ^2 \left(2 {y_{1c}}^4-3
   {y_{1c}}^2+1\right)-6\right)}{6 \lambda ^2 {y_{1c}}^2 \left(1-2 {y_{1c}}^2\right)^2}+{u_2}  + \mathcal{O}(3),  \label{center1a}\\
   & u_2'=-\frac{{u_1} {u_2}^2}{4 {y_{1c}}^6-6 {y_{1c}}^4+2 {y_{1c}}^2}+\frac{{u_2}^3 \left(\lambda ^2
   {y_{1c}}^4+3\right)}{6 {y_{1c}}^4 \left({y_{1c}}^2-1\right) \left(\lambda -2 \lambda  {y_{1c}}^2\right)^2}-\frac{{u_2}^2}{4 {y_{1c}}^4-6 {y_{1c}}^2+2} + \mathcal{O}(3) \label{center1b}. 
\end{align}

\begin{figure}[ptb]
\centering\includegraphics[width=0.8\textwidth]{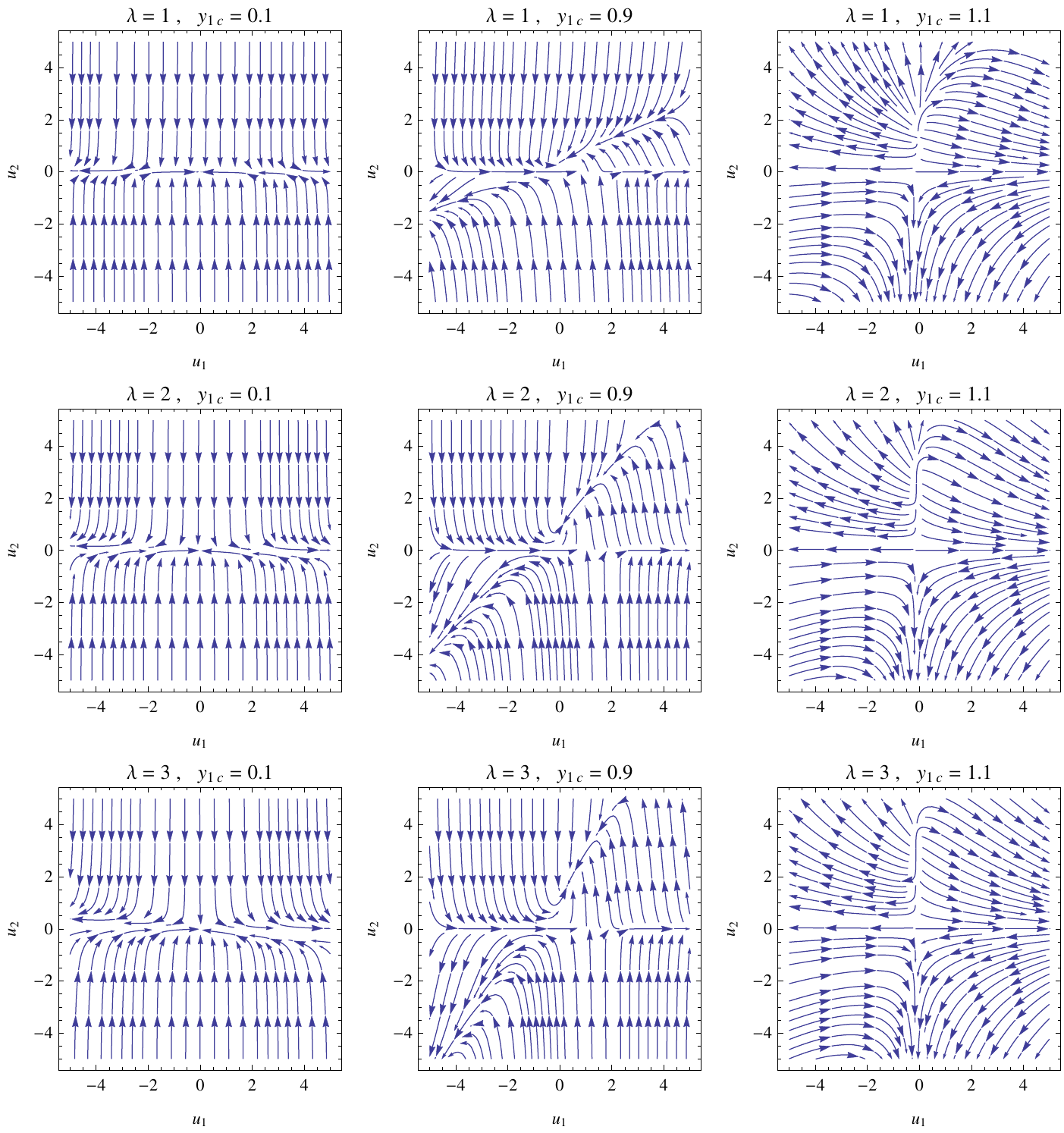} \caption{Dynamics on the center manifold is given locally (up to second order) by \eqref{center1a}, \eqref{center1b} for several choices of the parameters. }%
\label{fig3}%
\end{figure}
Defining the function 
\begin{equation}
    U(u_1, u_2)=\frac{{u_1}^2}{2}+{u_2}^2, 
\end{equation}
we have 
\begin{align}
   & U'(\tau)=\nabla U \cdot (u_1', u_2')=u_1 u_2 + \frac{{u_1}^2 {u_2} \left(2 {y_{1c}}^6+7 {y_{1c}}^4-8 {y_{1c}}^2+1\right)}{4 {y_{1c}}^6-6 {y_{1c}}^4+2 {y_{1c}}^2} \nonumber \\
    &+{u_1} \left(\frac{{u_2}^2 \left(\lambda ^2 \left(2 {y_{1c}}^4-3
   {y_{1c}}^2+1\right)-6\right)}{6 \lambda ^2 {y_{1c}}^2 \left(1-2 {y_{1c}}^2\right)^2}-\frac{{u_2}^3}{2 {y_{1c}}^6-3 {y_{1c}}^4+{y_{1c}}^2}\right)\nonumber \\
   & +\frac{{u_2}^4 \left(2 \lambda ^2 {y_{1c}}^4+6\right)}{6
   {y_{1c}}^4 \left({y_{1c}}^2-1\right) \left(\lambda -2 \lambda  {y_{1c}}^2\right)^2}+\frac{{u_2}^3}{-2 {y_{1c}}^4+3 {y_{1c}}^2-1}\sim u_1 u_2 + \mathcal{O}(3).
\end{align}
Hence, for small $u_1, u_2$ such as $u_1 u_2 < 0$ the origin is locally stable, whereas,  for small $u_1, u_2$ such as $u_1 u_2 >0$ it is unstable.
\begin{figure}[ptb]
\centering\includegraphics[width=0.8\textwidth]{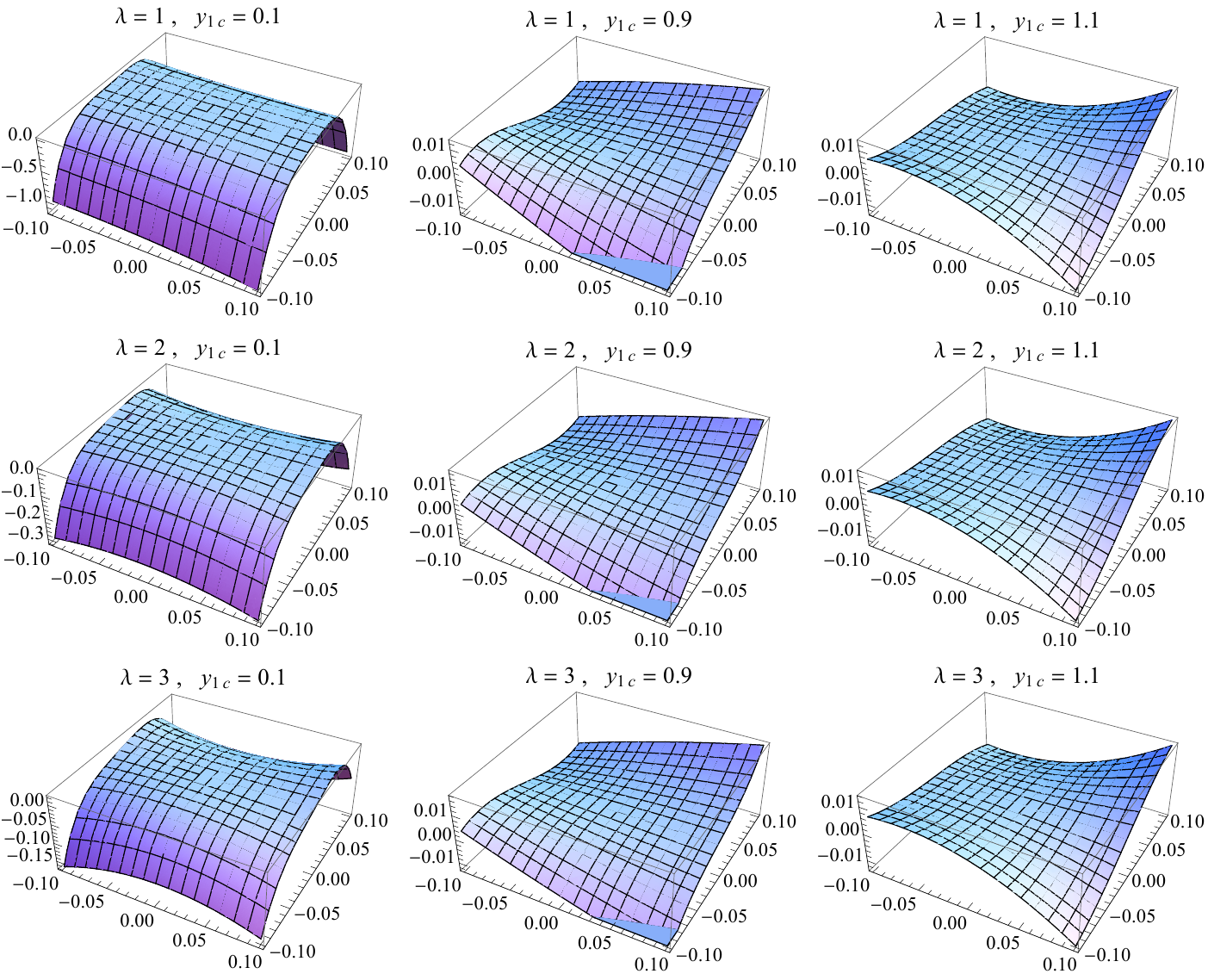} \caption{$U'(\tau)$ for several choices of the parameters, showing that for small $u_1, u_2$ such as $u_1 u_2 < 0$ the origin is locally stable, whereas,  for small $u_1, u_2$ such as $u_1 u_2 >0$ it is unstable.}%
\label{fig4}%
\end{figure}

In the figure \ref{fig3} is it presented the dynamics on the center manifold is given locally (up to second order) by \eqref{center1a}, \eqref{center1b} for several choices of the parameters.
In figure \ref{fig4} it is presented  $U'(\tau)$ for the same choices of the parameters, showing that for small $u_1, u_2$ such as $u_1 u_2 < 0$ the origin is locally stable, whereas,  for small $u_1, u_2$ such as $u_1 u_2 >0$ it is unstable.

\begin{figure}[ptb]
\centering
\includegraphics[width=0.3\textwidth]{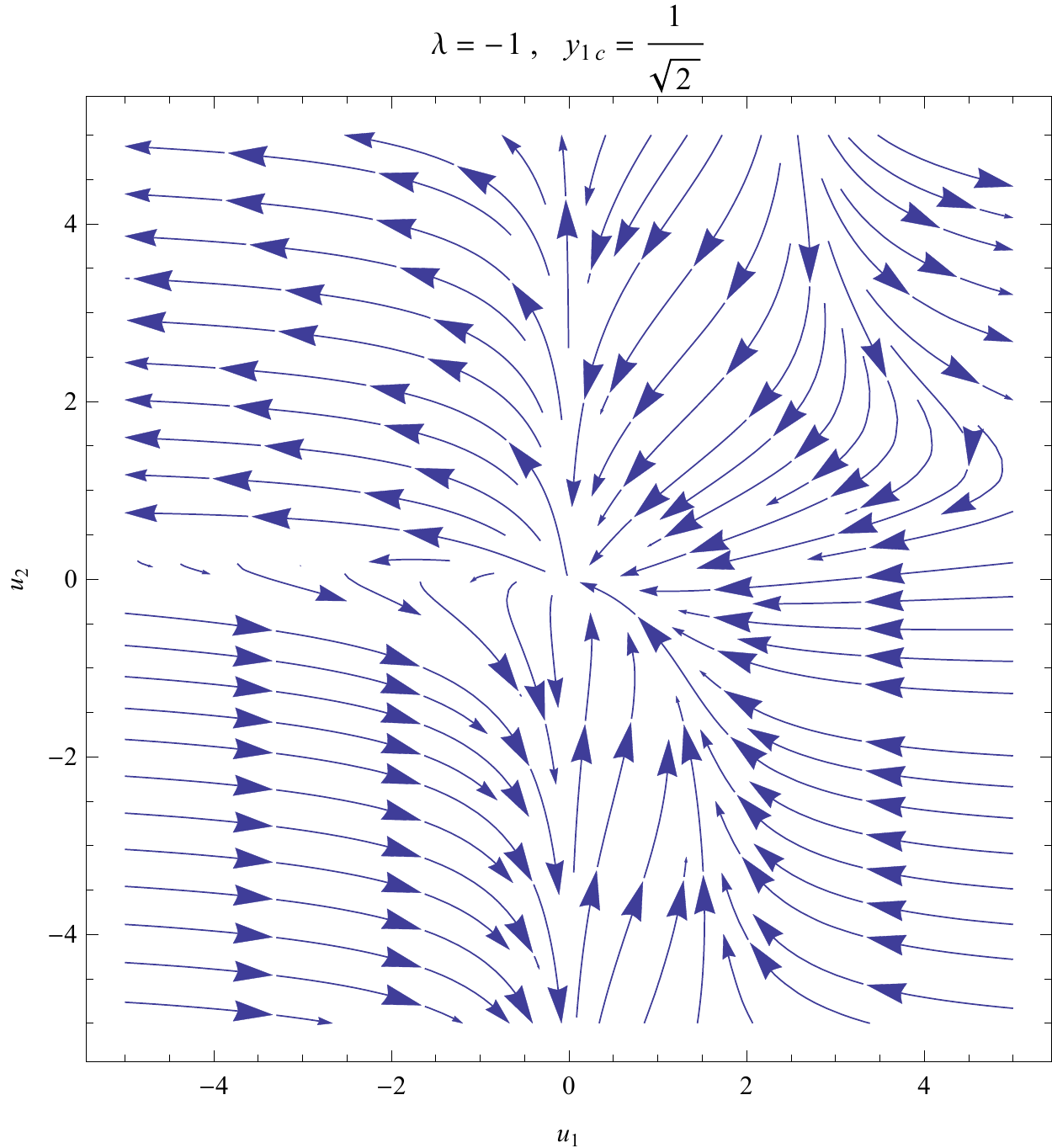} 
\includegraphics[width=0.3\textwidth]{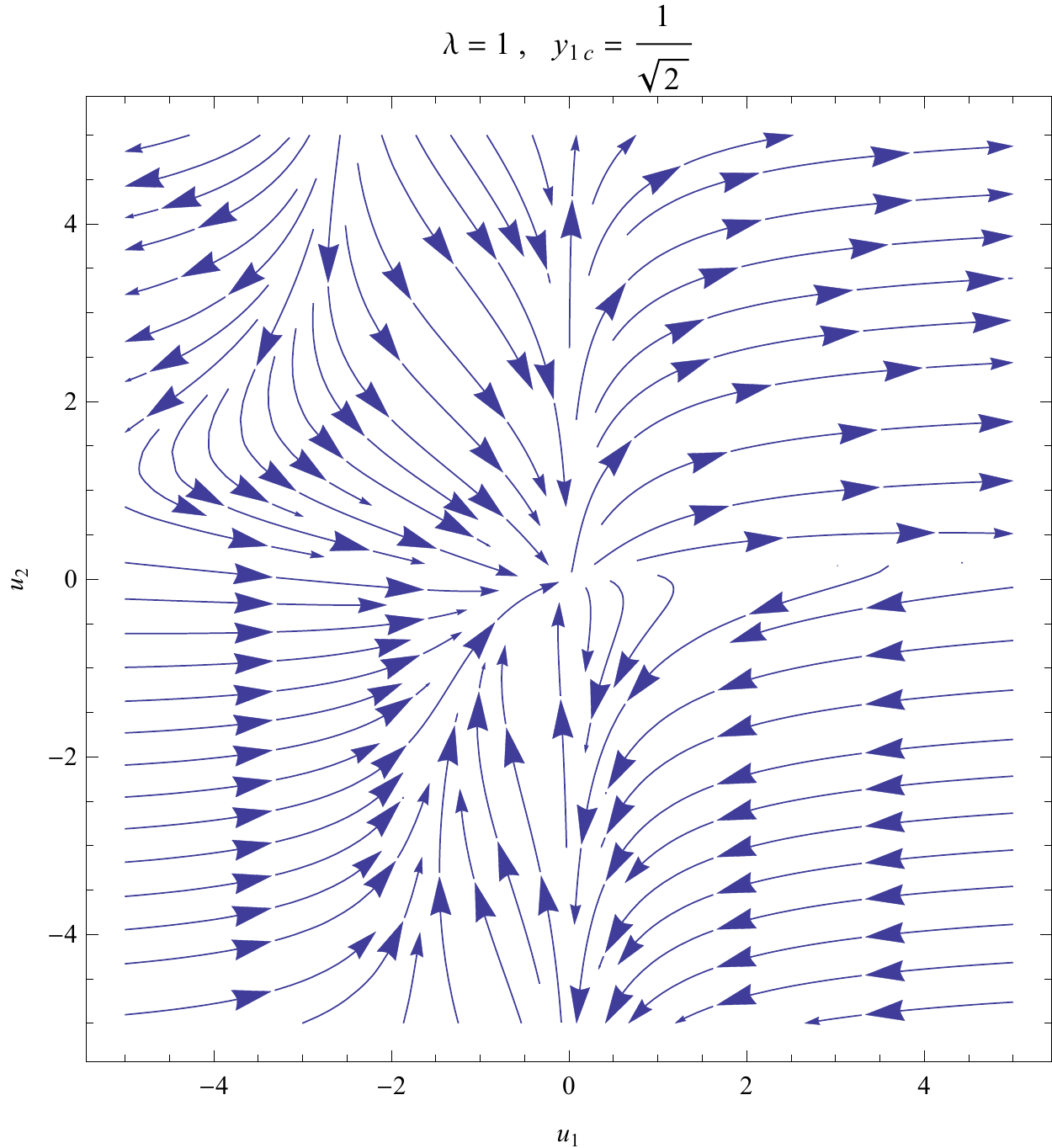} \caption{Dynamics on the center manifold is given locally (up to second order) by \eqref{center3a}, \eqref{center3b} for $\lambda=-1$, (resp. $\lambda=1$). The dynamics is qualitative the same for other values of $\lambda<0$ (resp. $\lambda>0$).}%
\label{center3}%
\end{figure}

\textbf{Special case $y_{1c}=\frac{\sqrt{2}}{2}$}. Using the transformation 
\begin{align}
 & {u_1}= \mu , \\
 & {u_2}= \frac{1}{24} \left(\lambda  \mu -2 \sqrt{6} \lambda  {x_1}+12 \sqrt{2} {y_1}+12 {y_2}-6\right),\\
 & {v_1}= \frac{1}{144} \left(-\sqrt{6} \left(\lambda ^2-12\right) \mu -12
   \lambda  \left(-2 \lambda  {x_1}+4 \sqrt{3} {y_1}+2 \sqrt{6} {y_2}+\sqrt{6}\right)\right)+{x_2},\\
 & {v_2}= -\frac{1}{48} \lambda  \left(\sqrt{6} (\lambda  \mu +18)-36 \lambda  {x_1}-24 \sqrt{3}
   {y_1}+12 \sqrt{6} {y_2}\right),\\
 & {v_3}= \frac{1}{24} \left(-\lambda  \mu +2 \sqrt{6} \lambda  {x_1}-12 \sqrt{2} {y_1}+12 {y_2}+18\right),
\end{align}
the linearization  transforms to 
\begin{equation}
 \left(
\begin{array}{c}
 v_1'\\
 v_2'\\
v_3'\\
u_1'\\
u_2'\\
\end{array}
\right) =   \left(
\begin{array}{ccccc}
 -3 & 0 & 0 & 0 & 0 \\
 0 & -3 & 1 & 0 & 0 \\
 0 & 0 & -3 & 0 & 0 \\
 0 & 0 & 0 & 0 & 0 \\
 0 & 0 & 0 & 0 & 0 \\
\end{array}
\right)\left(
\begin{array}{c}
 v_1\\
 v_2\\
v_3\\
u_1\\
u_2\\
\end{array}
\right).
\end{equation}
Applying the CMT we obtain that the center manifold is given locally (up to second order) by the graph
\begin{align}
   \Bigg\{(u_1,u_2, v_1, v_2, v_3)\in\mathbb{R}^5: \quad   & v_1 = a_1 u_1^2 + a_2 u_1 u_2 + a_3 u_2^2 + \mathcal{O}(3), \nonumber \\ 
    & v_2 = b_1 u_1^2 + b_2 u_1 u_2 + b_3 u_2^2 + \mathcal{O}(3), \nonumber \\  
    & v_3 = c_1 u_1^2 + c_2 u_1 u_2 + c_3 u_2^2 + \mathcal{O}(3)\Bigg\},
\end{align}
where
\begin{align}
& a_1= 0, a_2= \frac{1}{\sqrt{6}}, a_3= 0, b_1=0, b_2= 0, b_3= 0, c_1=-\frac{1}{192}, c_2= 0, c_3= \frac{1}{4}.
\end{align}
Hence, the dynamics on the center manifold is given locally (up to second order) by 
\begin{align}
   &u_1'= \frac{1}{192} {u_1}^2 \left({u_1} \left(-\lambda ^2+24 {u_2}-12\right)+24 (\lambda +2 \lambda  {u_2})\right) + \mathcal{O}(3),  \label{center3a}\\
   & u_2'=-\frac{1}{32} {u_1} ({u_1}-8 \lambda  {u_2}) + \mathcal{O}(3) \label{center3b}. 
\end{align}
According to Figure \ref{center3}, the origin is a saddle. 

\begin{figure}[ptb]
\centering
\includegraphics[width=0.3\textwidth]{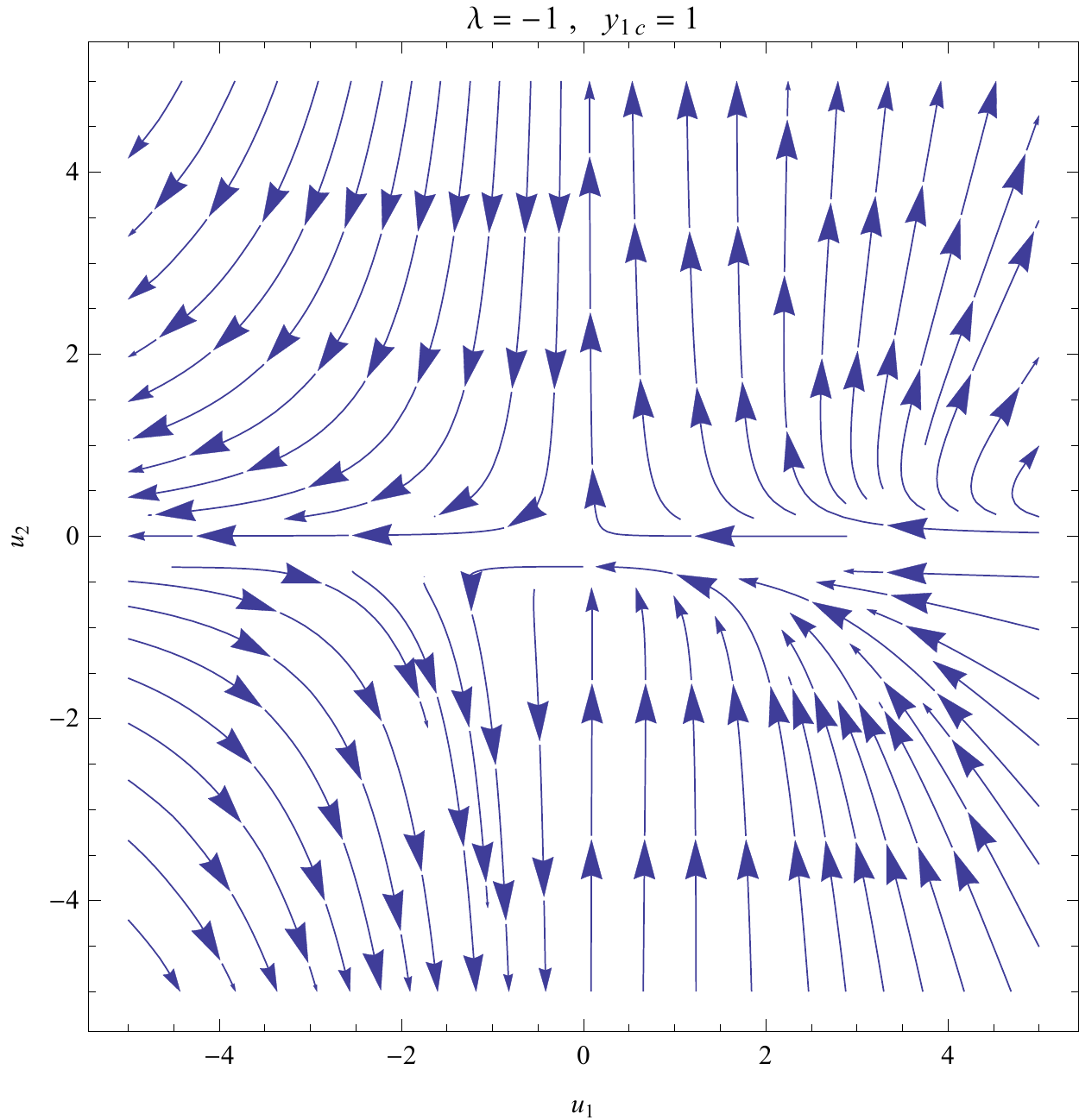} 
\includegraphics[width=0.3\textwidth]{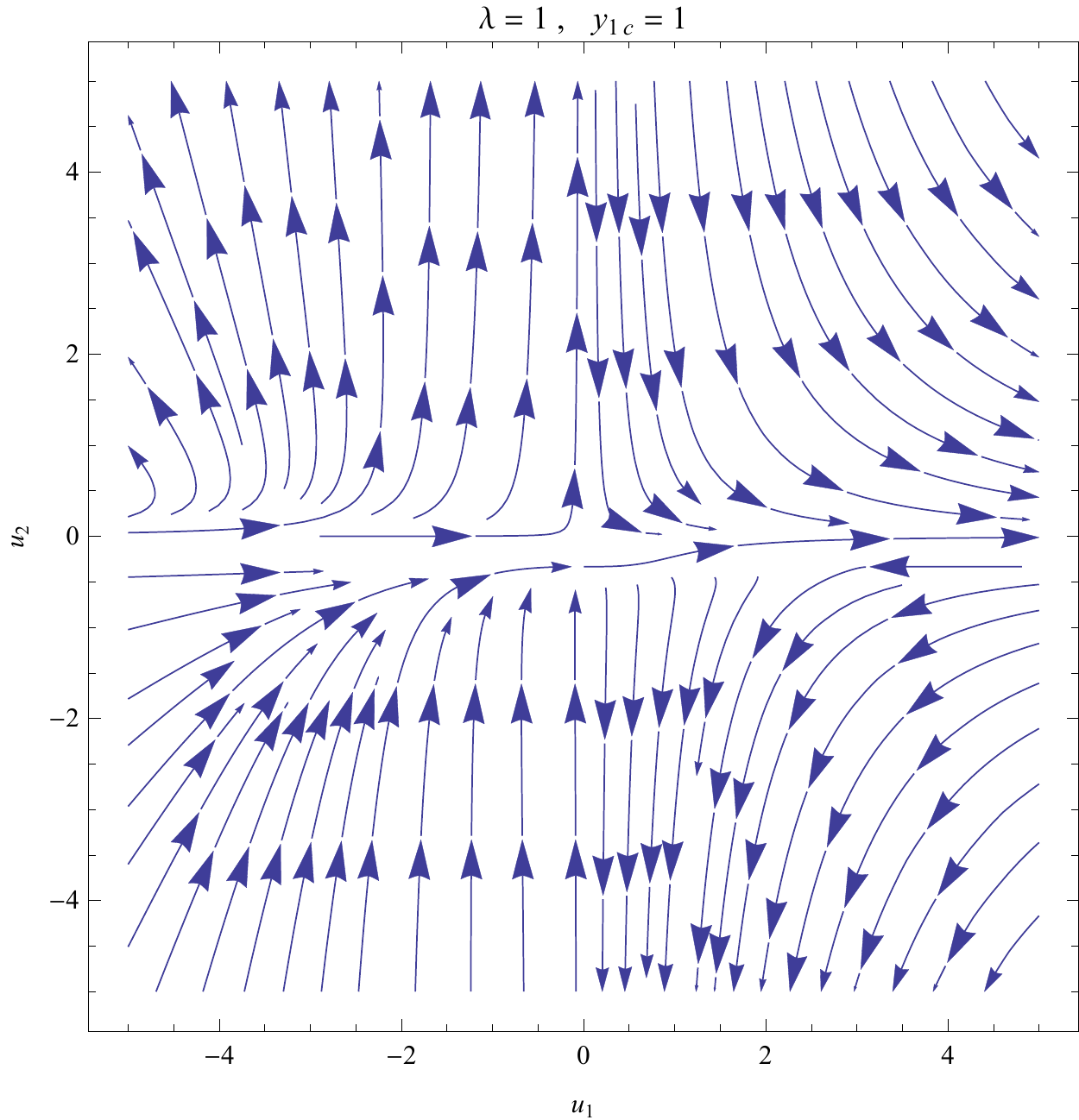} \caption{Dynamics on the center manifold is given locally (up to second order) by \eqref{center4a}, \eqref{center4b} for $\lambda=-1$, (resp. $\lambda=1$). The dynamics is qualitative the same for other values of $\lambda<0$ (resp. $\lambda>0$).}%
\label{center4}%
\end{figure}

\textbf{Special case $y_{1c}=1$}. Using the transformation 
\begin{equation}
    u_1= \mu ,u_2= y_2, v_1= \frac{1}{2} \left(-2 \sqrt{6} \lambda +2 \lambda ^2 x_1+2 \sqrt{6} \lambda  y_1-\sqrt{6} \lambda  y_2\right), v_2= \frac{1}{2} (2
   y_1-y_2-2), v_3= \frac{1}{3} \left(-\sqrt{6} \lambda +3 x_2-\sqrt{6} \lambda  y_2\right),
\end{equation}
the linearization  transforms to 
\begin{equation}
 \left(
\begin{array}{c}
 v_1'\\
 v_2'\\
v_3'\\
u_1'\\
u_2'\\
\end{array}
\right)= \left(
\begin{array}{ccccc}
 -3 & 1 & 0 & 0 & 0 \\
 0 & -3 & 0 & 0 & 0 \\
 0 & 0 & -3 & 0 & 0 \\
 0 & 0 & 0 & 0 & 0 \\
 0 & 0 & 0 & 0 & 0 \\
\end{array}
\right)\left(
\begin{array}{c}
 v_1\\
 v_2\\
v_3\\
u_1\\
u_2\\
\end{array}
\right).
\end{equation}
Applying the CMT we obtain that the center manifold is given locally (up to second order) by the graph
\begin{align}
   \Bigg\{(u_1,u_2, v_1, v_2, v_3)\in\mathbb{R}^5: \quad   & v_1 = a_1 u_1^2 + a_2 u_1 u_2 + a_3 u_2^2 + \mathcal{O}(3), \nonumber \\ 
    & v_2 = b_1 u_1^2 + b_2 u_1 u_2 + b_3 u_2^2 + \mathcal{O}(3), \nonumber \\  
    & v_3 = c_1 u_1^2 + c_2 u_1 u_2 + c_3 u_2^2 + \mathcal{O}(3)\Bigg\},
\end{align}
where
\begin{align}
& a_1=0, a_2=0, a_3=-\frac{1}{4} \sqrt{\frac{3}{2}} \lambda , b_1=0, b_2= \frac{\lambda }{12}, b_3= -\frac{1}{8}, c_1=0, c_2=\frac{\lambda ^2+3}{3 \sqrt{6}}, c_3=0.
\end{align}
Hence, the dynamics on the center manifold is given locally (up to second order) by 
\begin{align}
   &u_1'=\frac{1}{24} u_1^2 \left(\left(\lambda ^2+3\right) u_1 u_2+6 \lambda  (u_2+1)\right)  + \mathcal{O}(3),  \label{center4a}\\
   & u_2'=-\frac{1}{2} \lambda  u_1 u_2 (3 u_2+1) + \mathcal{O}(3) \label{center4b}. 
\end{align}
According to Figure \ref{center4}, the origin is a saddle. 

\subsection{Equilibrium points for arbitrary potential}

Consider now an arbitrary potential $V\left(  \phi\right)  $, such that
$\Gamma\left(  \lambda\right)  $ to be arbitrary. In that consideration, the
dynamical system of our consideration has dimension six, where the equilibrium
points have coordinates $\bar{P}=\left(  x_{1}\left(  \bar{P}\right)
,x_{2}\left(  \bar{P}\right)  ,y_{1}\left(  \bar{P}\right)  ,y_{2}\left(
\bar{P}\right)  ,\mu\left(  \bar{P}\right)  ,\lambda\left(  \bar{P}\right)
\right)  $. 

In particular the equilibrium points are $\bar{P}_{A}=\left(  x_{1}\left(
P_{A}\right)  ,x_{2}\left(  P_{A}\right)  ,y_{1}\left(  P_{A}\right)
,y_{2}\left(  P_{A}\right)  ,\mu\left(  P_{A}\right)  ,\lambda\left(
P_{A}\right)  \right)  $, where $A=1,2,3$ and $\lambda\left(  P_{A}\right)  $
solves the algebraic equation $\lambda^{2}\left(  \Gamma\left(  \lambda
\right)  -1\right)  =0$. While there are more equilibrium points because of the
different values of $\lambda\left(  P_{A}\right)  $, the physical physical
properties of the equilibrium points are exactly the same with the exponential
potential. That is an interesting result, since any kind of potential has a
similar physical behaviour. 

Since the physical properties on the additional equilibrium points do not change, we omit the presentation of the stability analysis. 

\section{Evolution of the Cosmographic parameters}
\label{sec6}

There are usually two known ways through which one tries to understand the dynamical history of our universe. One is the model independent approach in which one deals with the kinematical quantities that are 
extracted  directly from the space-time metric and secondly, one needs to fix either the matter sector of the universe or the underlying gravitational sector and then explore the dynamics of the universe. 
Indeed, the kinematical quantities are more flexible and novel compared to the model dependent approach since within the kinematical approach the dynamics of the universe is solely dependent on the geometrical quantities and hence this is more robust.  The study of the kinematical quantities derived from a homogeneous and isotropic universe is called the cosmography \cite{Weinberg-GR}. Using the variation of the kinematical quantities, one can measure the viability of a proposed cosmological scenario.

Under the background of the homogeneous and isotropic universe characterized by the FLRW universe, one can define the kinematical quantities as  \cite{Visser:2003vq,Bolotin:2018xtq}

\begin{eqnarray}
	&& H(t)  = \frac1a\frac{da}{dt}, \nonumber \\
	&&q(t) =  -\frac1a\frac{d^2a}{dt^2}\left[\frac1a\frac{da}{dt}\right]^{-2}, \nonumber \\
	&&j(t) =  \frac1a\frac{d^3a}{dt^3}\left[\frac1a\frac{da}{dt}\right]^{-3},\nonumber \\
	&&s(t) =  \frac1a\frac{d^4a}{dt^4}\left[\frac1a\frac{da}{dt}\right]^{-4},\nonumber 
\end{eqnarray}
where $H$, $q$, $j$, $s$  are respectively called the Hubble, deceleration, jerk, snap  parameters and except for the Hubble parameter, the last three parameters are dimensionless. With these terminology, the expansion scale factor $a(t)$ can be written as 

\begin{eqnarray}
    a(t) =  a_0 \left[1 + H_0 (t-t_0) - \frac{1}{2} q_0^2 (t-t_0)^2 + \frac{1}{3!} j_0 (t-t_0)^3 + \frac{1}{4!} s_0 (t-t_0)^4 + \mathcal{O} [(t-t_0)^5] \right],
\end{eqnarray}
in which any sub-index attached to any quantity refers to its present value. 
We note that one can extend this series by allowing higher order derivatives of the scale factor beyond order four. Here we are interested to investigate the deceleration parameter, and its next two hierarchies, namely, jerk and the snap parameter which in terms of the Hubble parameter can alternatively be rewritten as 

\begin{align}
q  & =-1-\frac{\dot{H}}{H^{2}}\\
j  & =\frac{\ddot{H}}{H^{3}}-3q-2\\
s  & =\frac{\dddot{H}}{H^{4}}+4j+3q\left(  q+4\right)  +6,
\end{align}
where an overhead dot represents the cosmic time derivative; therefore, two or three overhead dots represents the two- or three-times derivative with respect to the cosmic time.   
Thus, from the evolution of $q, j, s$, one can understand the expansion of the universe (accelerating or decelerating), the rate of acceleration ($j$) and its next derivative. 
In Fig. \ref{fig3} we present the qualitative evolution of the cosmographic  parameters. 

\begin{figure}[ptb]
\centering\includegraphics[width=0.8\textwidth]{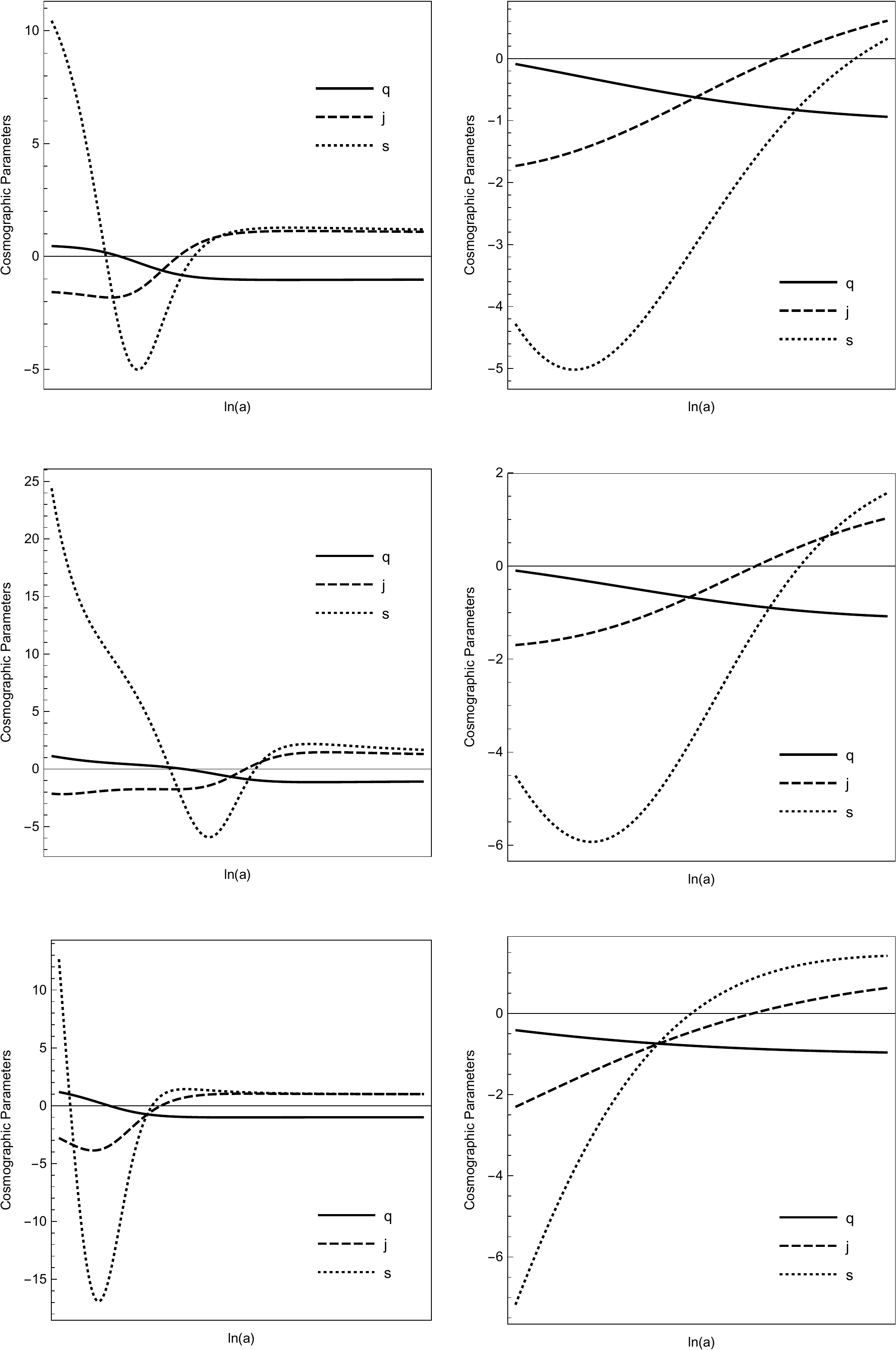} \caption{Qualitative
evolution of the cosmographic parameters $q,~j$ and $s$, for the same initial  conditions as used for Fig. \ref{fig1}. Right Figs are subplots of the Left Figs in the  range near the de Sitter attractor. }%
\label{fig3}%
\end{figure}

\section{Conclusions}
\label{sec7}

The dynamics of the universe both at its very early and late evolution are two mysterious issues in cosmology.  Various approaches have been proposed to understand both the phases of the universe but none of them are found to be extremely satisfactory.  The classical theory of gravity does not work out during the very early evolution of the universe but the quantum gravity does a bit, according to the available quantum gravitational theories. The quantum gravity theory predicts 
the existence of a minimum length scale of the order of the Planck length $l_{\rm pl}$ and consequently, this motivated to generalize  the
Heisenberg's Uncertainty Principle to some Generalized Uncertainty principle (GUP) in quantum gravity.  
The Generalized Uncertainty principle has been found to be very effective to explain several important issues that are directly related to our universe and its evolution. In particular, in the context of black hole evaporation, GUP has a significant role. Moreover, GUP has a explanation towards the existence of the dark matter fluid in the universe's sector. Thus, it will be interesting whether the GUP modified cosmological models could give rise to some interesting features of the universe.  
Hence, in the present work we study the modified quintessence model and investigate the stability of the model by performing its dynamical analysis.

We first investigate the modified quintessence model for the exponential potential which gives rise to various possibilities. For various critical points,  we get different solutions such as the matter dominated universe and the universe dominated by the kinetic term only, however, all of them are found to be unstable in nature. Only the family of points on the surface with coordinates $P_3$ describe the de Sitter universes and they all are attractors. It is interesting to note that although no scaling solutions or a tracking solution exist in this specific model contrary to the quintessence
theory, however, the critical points describing the de Sitter  solution do not exist in the case of a quintessence model with an
the exponential potential.  This is a new result in this work. In order to understand the behavior of the family of critical points, $P_3$, we applied the Center Manifold Theorem.

Furthemrore, we considered an arbitrary potential and performed the dynamical system analysis similar to the exponential potential. We found that the  physical properties of the equilibrium points for the general potential remain exactly same as in the case with the exponential potential. That is a very striking and exciting result in this work, because this clearly indicates that the investigations with the exponential potential is enough for the modified quintessence models with GUP.  

Finally, we observe that the parameter for the equation of state for the effective cosmological fluid can cross the phantom divine line and as it was observed by the numerical simulations without the appearance of ghosts, while the late time attractor is the de Sitter universe.

\begin{acknowledgments}
AG was funded by FONDECYT grant number 1200293. The research of AP and GL was funded by Agencia Nacional de Investigaci\'{o}n y
Desarrollo - ANID through the program FONDECYT Iniciaci\'{o}n grant no.
11180126. Additionally, by Vicerrector\'{\i}a de Investigaci\'{o}n y
Desarrollo Tecnol\'{o}gico at Universidad Catolica del Norte. SP was supported
by the  Mathematical Research Impact-Centric Support Scheme (MATRICS), File No.
MTR/2018/000940, awarded by the Science and Engineering Research Board (SERB),  Govt. of India.
\end{acknowledgments}


\begin{thebibliography}{9}                                                                                              %
\bibitem {guth}
A. Guth, Phys. Rev. D {\bf 23}, 347 (1981).


\bibitem{linde}
A. Linde, 
Phys. Lett. B {\bf 108}, 389 (1982).

\bibitem{copeland}
E. J. Copeland, M. Sami and S. Tsujikawa,  Int. J. Mod. Phys. D {\bf 15}, 1753 (2006).


  \bibitem{Clifton1} 
 T. Cclifton, P.G. Ferreira, A. Padilla and C. Skordis, Phys. Rept. \textbf{513}, 1 (2012)

\bibitem{Nojiri:2017ncd} 
  S.~Nojiri, S.~D.~Odintsov and V.~K.~Oikonomou,
  Phys.\ Rept.\  {\bf 692}, 1 (2017).
  
  


\bibitem{MB01} J. Martin and R. H. Brandenberger, Phys. Rev. D {\bf 63}, 123501 (2001).
\bibitem{Niem01} J. C. Niemeyer, Phys. Rev. D {\bf 63}, 123502 (2001).
\bibitem{Kowa01} J. Kowalski-Glikman, Phys. Lett. B {\bf 499},1 (2001).
\bibitem{Niem01b} J. C. Niemeyer and R. Parentani Phys. Rev. D {\bf 64} 101301 (2001).
\bibitem{Kempf01} A. Kempf, Phys. Rev. D {\bf 63}, 083514 (2001), astro-ph/0009209.
\bibitem{KN01} A. Kempf and J. Niemeyer, Phys. Rev. D {\bf 64}, 103501 (2001).
\bibitem{EGKS01} R. Easther, B. Greene, W. H. Kinney and G. Shiu, Phys. Rev. D {\bf 64}, 103502 (2001).

\bibitem{Amjad1} A. Ashoorioon, A. Kempf and R.B. Mann, Phys. Rev. D \textbf{71}, 023503 (2005).

\bibitem{Amjad2} A. Ashoorioon, J. L. Hovdebo and R. B. Mann, Nucl. Phys. B \textbf{727}, 63-76 (2005).


\bibitem{Mukhi:2011zz}
S.~Mukhi,
Class. Quant. Grav. \textbf{28}, 153001 (2011).  

\bibitem{KowalskiGlikman:2004qa}
J.~Kowalski-Glikman,
Lect. Notes Phys. \textbf{669}, 131-159 (2005). 



\bibitem{AmelinoCamelia:2010pd}
G.~Amelino-Camelia,
Symmetry \textbf{2}, 230-271 (2010). 

\bibitem{Ghosh:2006bx}
S.~Ghosh,
Phys. Lett. B \textbf{648}, 262-265 (2007). 

\bibitem{Pramanik:2012fj}
S.~Pramanik, S.~Ghosh and P.~Pal,
Annals Phys. \textbf{346}, 113-128 (2014)

\bibitem{Bekenstein1} J. D. Bekenstein, Phys. Rev. D \textbf{7}, 2333 (1973). 
\bibitem{Bekenstein2} J. D. Bekenstein, Letter Nuovo Cimento \textbf{4}, 737 (1972). 

\bibitem{Bekenstein3}
J.~D.~Bekenstein,
Phys. Rev. D \textbf{9}, 3292-3300 (1974). 

\bibitem{Bekenstein4} J. D. Bekenstein, Stud. Hist. Phil. Sci. B \textbf{32}, 511-524 (2001). 


\bibitem {Maggiore}M. Maggiore, Phys. Lett. B \textbf{304}, 65 (1993).  

\bibitem{SFW86} Y. Sofue, M. Fujimoto and R.Wielebinski, Ann. Rev. Astron. Astrophys. \textbf{24}, 459 (1986).

\bibitem{K94} P. P. Kronberg, Rep. Prog. Phys. \textbf{57}, 325 (1994).

\bibitem{Amjad04} Amjad Ashoorioon and Robert B. Mann, Phys. Rev. D \textbf{71}, 103509 (2005). 



\bibitem{Hawking74} S. W. Hawking, Comm. Math. Phys. \textbf{43}, 199 (1974); 
\bibitem{BD82} N. D. Birrell and P.C. W. Davies, Quantum Fields in Curved Space, Cambridge Univ. Press, (1982).


\bibitem{Bunch81} T. S. Bunch, J. Phys. A: Math Gen. \textbf{14}, L139 (1981).
\bibitem{CMP88} C. G. Callan, R. C. Myers, and M. J. Perry, Nucl. Phys. B \textbf{311}, 673-698 (1989). 

\bibitem{York} J. W. York Jr, Quantum Theory of Gravity, Hilger 1984.

\bibitem{Parikh99} M. K. Parikh and F. Wilczek, Phys. Rev. Lett. \textbf{85}, 5042-5045 (2000). 

\bibitem{Susskind95} L. Susskind, J. Math. Phys. \textbf{36}, 6377 (1995).

\bibitem{Adler} Ronald J. Adler, P. Chen and D. I. Santiago, Gen. Rel. Grav. \textbf{33}, 2101-2108 (2001). 

\bibitem{Cline} D. B. Cline, Phys. Rep. \textbf{307}, 1 (1998).


\bibitem {Quesne2006}C. Quesne and V.M. Tkachuk, J. Phys. A \textbf{39}, 10909 (2006).

\bibitem {Vagenas}S. Das and E.C. Vagenas, Phys. Rev. Lett. \textbf{101}, 221301 (2008).

\bibitem {Kemph1}A. Kempf, J. Phys. A: Math. Gen. \textbf{30}, 2093 (1997).

\bibitem {Kemph2}H. Hinrichen and A. Kempf, J. Math. Phys. \textbf{37}, 2121 (1996).

\bibitem {ss1}S. Masood, M.\ Faizal, Z. Zal, A.F.\ Ali, J. Raza and M.B. Shah,
Phys. Lett. B \textbf{763}, 218 (2016)

\bibitem {Moayedi}S.K. Moayedi, M.R. Setare and H Moayeri, Int. J. Theor.
Phys. \textbf{49}, 2080 (2010).

\bibitem {ratra1}B. Ratra and P.J.E. Peebles, Phys. Rev.\ D \textbf{37}, 3406 (1988)

\bibitem {Saridakis1}E.N. Saridakis and J. M. Weller, Phys. Rev. D \textbf{81}, 123523 (2010).


\bibitem{Paliathanasis:2015cza}
A.~Paliathanasis, S.~Pan and S.~Pramanik,
Class. Quant. Grav. \textbf{32}, no.24, 245006 (2015)

\bibitem{Weinberg-GR}  S. Weinberg, Gravitation  and  cosmology:  Principles  and  applications  of the  general  theory  of relativity, (Wiley, New York, 1972).

\bibitem{Visser:2003vq}
M.~Visser,
Class. Quant. Grav. \textbf{21}, 2603-2616 (2004).

\bibitem{Bolotin:2018xtq}
Y.~L.~Bolotin, V.~A.~Cherkaskiy, O.~Y.~Ivashtenko, M.~I.~Konchatnyi and L.~G.~Zazunov,
[arXiv:1812.02394 [gr-qc]].



\end{thebibliography}
\end{document}